# Reconfigurable entanglement distribution network based on pump management of spontaneous four-wave mixing source

Jingyuan Liu[1], Dongning Liu[1], Zhanping Jin[1], Zhihao Lin[1], Hao Li[2], Lixing You[2], Xue Feng[1], Fang Liu[1], Kaiyu Cui[1], Yidong Huang[1, 3] and Wei Zhang[1, 3,*]

[1] Frontier Science Center for Quantum Information, Beijing National Research Center for Information Science and Technology (BNRist), Electronic Engineering Department, Tsinghua University, Beijing 100084, China.

[2] National Key Laboratory of Materials for Integrated Circuits, Shanghai Institute of Microsystem and Information Technology, Chinese Academy of Sciences, Shanghai 200050, China

[3] Beijing Academy of Quantum Information Sciences, Beijing 100193, China.

*Corresponding author. E-mail address: zwei@tsinghua.edu.cn.*

**Abstract:**

Leveraging the unique properties of quantum entanglement, quantum entanglement distribution networks support multiple quantum information applications and are essential to the development of quantum networks. However, its practical implementation poses significant challenges to network scalability and flexibility. In this work, we propose a novel reconfigurable entanglement distribution network based on tunable multi-pump excitation of a spontaneous four-wave mixing (SFWM) source and a time-sharing method. We characterize the two-photon correlation under different pump conditions to demonstrate the effect of pump degenerate and pump non-degenerate SFWM processes on the two-photon correlation, and its tunability. Then as a benchmark application, a 10-user fully-connected quantum key distribution (QKD) network is established in a time-sharing way with triple pump lights. Each user receives one frequency channel thus it shows a linear scaling between the number of frequency channels and the user number in despite of the network topology. Our results thus provide a promising networking scheme for large-scale entanglement distribution networks owing to its scalability, functionality, and reconfigurability.

## 1 Introduction

Quantum entanglement distribution network is an important stage in the development of quantum networks, which generates end-to-end quantum entanglement in a deterministic or heralded way[1]. It enables various quantum information applications, such as quantum key distribution (QKD)[2-4], quantum teleportation[5, 6], distributed quantum computing[7], and quantum metrology and sensing[8, 9]. The most intriguing thing among these applications is the potential to implement device-independent protocols[10, 11]. To realize entanglement distribution between multiple users, a natural method is using wavelength division multiplexing (WDM). So far, based on WDM, efforts have been made to realize entanglement distribution networks via active routing by optical switches[12, 13] or passive splitting by beam splitters[14]. However, the network topologies of these works are simple and they suffer from problems caused by the duty cycle of optical switches or probabilistic splitting. In addition, the entanglement distribution network with trusted nodes[15] was proposed to increase the user number of a QKD network, while it still has potential security risks due to these nodes.

In recent years, there have been several remarkable works exploring the construction method of entanglement distribution networks. The concept of a fully-connected network based on WDM is proposed and implemented with 4 users[16]. In this scheme, each user pair exclusively occupies a pair of frequency channels. It is an ideal case for the entanglement distribution network because there is a one-to-one correspondence between the user link and the frequency pair of entanglement resource. However, the number of required wavelength channels is $O(N^2)$ for $N$ users, which limits the further increase of user number since the wavelength channels are precious resources. To increase the user number of a fully-connected network, passive multi-port beam splitters are used to probabilistically distribute an entanglement resource to multiple user pairs[17, 18]. This strategy is rather suitable for large-scale QKD networks with post-processing, while it may cause problems in other applications, such as teleporting a quantum state. Recently, some works introduced wavelength selective switch (WSS)[19-21] technology and the concept of a q-ROADM[22, 23], which consists of filters, optical fiber switches, and WSS, to the entanglement distribution network to dynamically reconfigure the network. Nevertheless, they also suffer from the problems of either limited user numbers or probabilistic entanglement distribution. Therefore, it is necessary to propose a networking scheme for entanglement distribution networks that can tackle these two problems simultaneously.

In this work, we present a reconfigurable entanglement distribution network architecture by using multiple tunable lasers to pump the spontaneous four-wave mixing (SFWM) process in a silicon waveguide chip, combined with a time-sharing method. We define this scheme as a pump management scheme, where the network topology is switched by tuning the pump frequencies and the physical structure of the network is unchanged. We first characterize the generated energy-time entangled photon pairs under different pump conditions. The quantum correlation between different frequency pairs shows a complex feature determined by the SFWM processes with degenerate and non-degenerate pump photons. The tunability of the two-photon correlation is demonstrated by changing the pump frequencies. Further, we demonstrate a 10-user fully-connected entanglement distribution network via pump management, and a QKD network is established as a benchmark application. The number of required wavelength channels is relaxed to $O(N)$ for $N$ users, rather than the quadratic relation in the WDM-based network. At the same time, there is no probabilistic beam splitting used in the network so that each user pair exclusively occupies a frequency pair of entanglement resources. It has the potential to support various quantum information applications beyond QKD. To the best of our knowledge, we have demonstrated the largest fully-connected quantum network without trusted nodes and probabilistic beam splitting to date, showing potential for full-fledged quantum networks in the future.

## 2 Methods

### 2.1 Pump management for SFWM

For $\chi^{(3)}$ nonlinear materials, SFWM enables the generation of energy-time entangled photon pairs, when energy conservation and phase matching conditions are satisfied[24]. The near-zero dispersion of the $\chi^{(3)}$ nonlinear materials, such as properly designed optical fibers and silicon wire waveguides, leads to the broadband characteristic of the entangled photon pairs. It also allows the frequency of the pump light to vary over a wide frequency range, adding an additional control parameter to manipulate the SFWM process. On the other hand, according to whether the two pump photons in an SFWM process have the same frequency, the SFWM processes are classified into two cases with degenerate pump photons and non-degenerate pump photons, as shown in Fig. 1(a). Compared with $\chi^{(2)}$ nonlinear processes, different energy conservation conditions of the SFWM processes provide the possibility to manipulate the spectral correlations between generated photon pairs by controlling the pump lights. For simplicity, the following discussion about degenerate and non-degenerate SFWM refers to the discussion about pump photons.

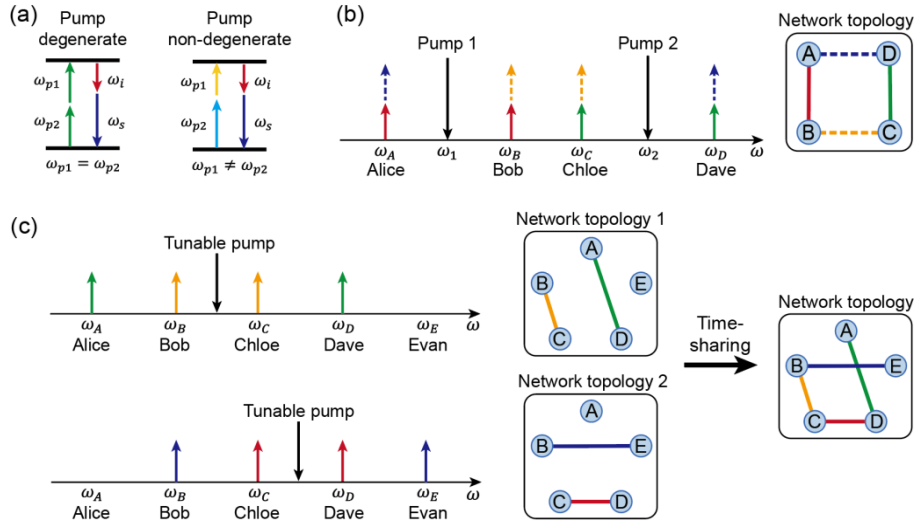

Fig 1. Schematics of entanglement distribution network using SFWM pump management. (a) Diagram of energy conservation relations for SFWM processes with degenerate and non-degenerate pump photons. (b) The pumping scheme of a ring-type quantum network via dual pumps. Two lasers with different frequencies are used to pump the $\chi^{(3)}$ nonlinear material. Solid arrows and dashed arrows represent photons generated by degenerate and non-degenerate SFWM processes, respectively. Here, a pair of entangled photons are shown in the same color. Using filters to select photons of different frequency channels and distribute them to different users respectively, then an entanglement distribution network with a specific topology is established. (c) The realization of a reconfigurable entanglement distribution network with a tunable pump. The frequency of the pump light is changed to realize different network topologies. Then the network is established in a time-

sharing way with more topological links.

Here, we define the pump management scheme as flexibly changing the number and frequency of pump lasers to realize a network with the desired topology. Through the management of SFWM pump lights, it is easy to realize a fully-connected entanglement distribution network in which each user receives photons in a single frequency channel. Firstly, in the case of multiple pump lights, several SFWM processes occur simultaneously, resulting in a complex two-photon correlation feature between photons with different frequencies. For instance, when two lasers pump the $\chi^{(3)}$ nonlinear material, both degenerate and non-degenerate SFWM processes will occur. The pumping scheme with dual pumps is shown in Fig. 1(b). The degenerate SFWM generates signal and idler photons with pairwise correlations that are symmetric about each pump frequency, shown by solid arrows. For the non-degenerate SFWM processes, the generated pairwise correlations are symmetric about the mean value of the two pump frequencies, shown by dashed arrows. In this case, if we select the specific four frequencies $\omega_A$, $\omega_B$, $\omega_C$ and $\omega_D$ with filters, and distribute them to four users respectively, a four-user entanglement distribution network with ring-type topology is established.

On the other hand, the network topology can be dynamically reconfigured by changing the frequency of pump lights. Using a time-sharing method, network topologies complement each other to form a network with more topological links, as shown in Fig. 1(c). The overall coincidence count rate is the total number of coincidence count events divided by the total time. The time-shared network reduces the number of user links that are connected simultaneously. Under some specific conditions, such as in systems with a high timing jitter of detector or low heralding efficiency of the entanglement source, it outperforms the network with all links connected at the same time[22, 23]. It is worth noting that each network user only receives photons in a single frequency channel, while the frequencies of pump lights change at different times. Therefore, the experimental setup will be quite simple compared with the WDM-based quantum networks.

## 2.2 Experimental setup

To demonstrate the effect of pump management on the generation of entangled photon pairs, we first characterize the two-photon correlation under different pump conditions. Fig. 2(a) illustrates the experimental setup to measure two-photon joint spectral intensity (JSI). Three continuous-wave (CW) pump lasers with tunable frequencies are multiplexed through standard dense wavelength division multiplexing (DWDM) filters. Then the light excites the SFWM process in a 10 mm long silicon waveguide. At the output, the pump light is rejected using DWDM filters. A programmable filter (Finisar Waveshaper) selects signal and idler photons with tunable frequencies and sends them to the superconducting nanowire single-photon detectors (SNSPDs). For the correlation measurement between non-degenerate signal and idler photons, two output ports of the Waveshaper are used. In the case of degenerate photons, only one output port of the Waveshaper is used, and a beam splitter (BS) further divides the photons into two parts before being detected by SNSPDs. Subsequently, the two-photon JSI is measured by time-resolved coincidence measurement of signal and idler photons with different frequencies.

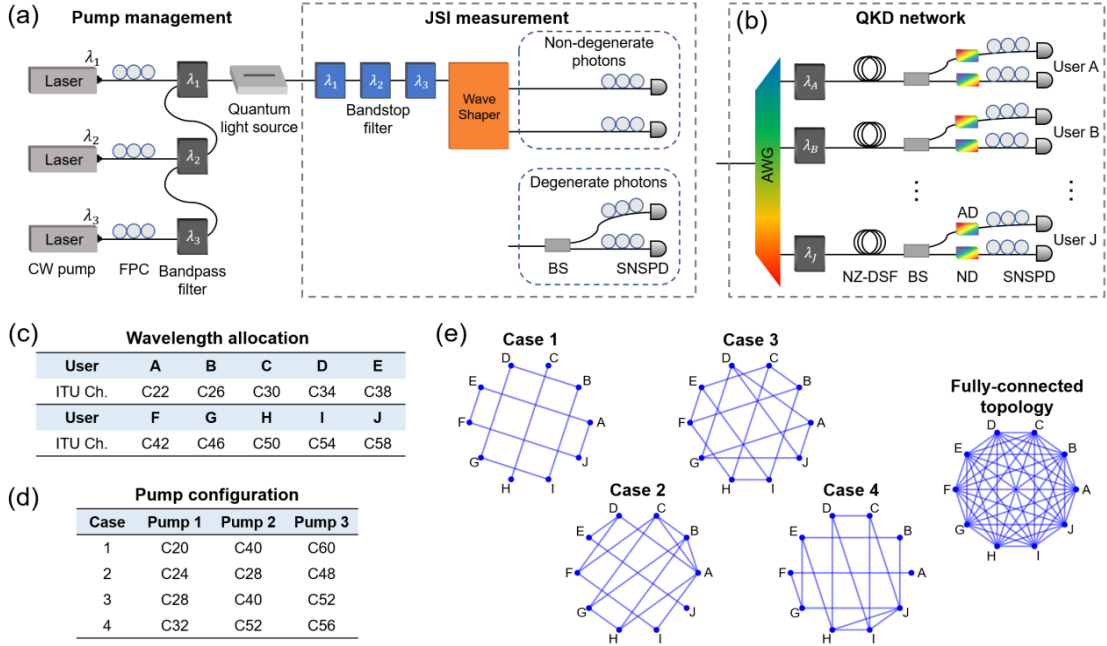

Fig. 2 Schematics of the experimental setup. (a) The setup for two-photon JSI measurement. The tunable pump lights are multiplexed using standard DWDM filters and coupled into a silicon waveguide chip to generate entangled photon pairs through SFWM. At the output, filters are used to reject the pump laser and select signal and idler photons with tunable frequencies. FPC: fiber polarization controller. (b) The setup for the 10-user QKD network after the quantum light source. The photons generated from the quantum light source are demultiplexed using AWG and DWDM filters. Each user receives photons in a single frequency channel. A beam splitter and dispersion modules with anomalous and normal dispersion are used to construct two detection bases of symmetric dispersion optics quantum key distribution (DO-QKD). AD: anomalous dispersion; ND: normal dispersion. (c) Wavelength allocation scheme and (d) pump configuration for the 10-user fully-connected entanglement distribution network. (e) Network topologies under four pump configurations and the generated fully-connected network topology using time-sharing.

By employing the pump management strategy, we implement a 10-user fully-connected entanglement-based QKD network, in which each user receives photons with a single frequency channel. The illustration of the QKD network experimental setup is schematically depicted in Fig. 2(b). The generated photons from the quantum light source are separated by an arrayed waveguide grating (AWG) system according to the 100-GHz International Telecommunication Union (ITU) channels. The noise photons outside the filter band are further suppressed using DWDM filters. Then, each user receives only one specific wavelength channel via a 6.2 km non-zero dispersion shifted fiber (NZ-DSF) spool. Therefore, the effective separation between every two users is 12.4 km.

For each user, a 50:50 beam splitter is used to randomly choose between the two detection bases. Then a symmetric dispersive optics QKD (DO-QKD) configuration[15] is constructed by placing the dispersion modules with normal and anomalous dispersion at two paths on each user's side. For a two-user QKD process, the two detection bases are defined as the cases where one user detects with normal dispersion and the other user detects with anomalous dispersion, and vice versa. Nonlocal dispersion cancellation[25] happens in each detection basis, which is the basis of security analysis[26-28]. The symmetric DO-QKD has been proven to be equivalent to the conventional DO-QKD scheme[26], while it is more suitable for entanglement-based QKD networks since every end user has the same receiver configuration. Another important feature of DO-QKD is that it encodes the arrival times of photons in a high-dimensional way to utilize single-photon events more efficiently. The details of DO-QKD key generation can be found in the Supplementary Material.

### 2.3 Network architecture

Unlike the WDM-based entanglement distribution network, whose topology is based on the distribution of frequency pairs, the topology of the network with pump management depends mostly on the pump scheme, rather than the physical experimental setup. In the process of designing the wavelength allocation and pump configuration schemes, three principles are mainly considered. First, in each pump configuration, the frequencies of pump lights and the users are separated by at least one ITU channel, to reduce the impact of residual pump photons on the performance. Second, the number of pump configurations to be switched is minimized to maximize the key rate in a time-sharing scheme. Third, the accumulated network topology is consistent with the desired topology, such as the fully-connected network in our scheme.

For a 10-user fully-connected QKD network, the users' wavelength allocation scheme is shown in Fig. 2(c). The 10 wavelength channels are evenly distributed in the telecom C-band. The pump frequencies are properly selected among the C-band ITU channels to avoid adjacent to the users' frequencies. If a pump frequency is the same as a user's frequency, this user will discard all photons in this situation. Based on the above principles, we design a 10-user fully-connected network with three pump excitations. The three pump lights switch between four configurations, as shown in Fig. 2(d). In each case, a QKD network is temporarily established with the designed topology shown in Fig. 2(e). The lines between different user nodes represent the quantum correlation of entangled photon pairs. Using the time-sharing method, a fully-connected network is established based on the results under the four pump configurations.

## 3 Results

### 3.1 Characterization of two-photon correlation under different pump conditions

To measure the two-photon correlation, we first pump the quantum light source using a single laser with a tunable frequency. The pumping scheme is depicted in Fig. 3(a) with the pump power before coupling into the chip of about 2 mW. For the JSI measurements in this work, the Waveshaper routes signal and idler photons whose central frequencies correspond to those of the 100-GHz ITU grid channels, however, the bandwidth is set to be 25 GHz to prevent a high single-photon count rate from saturating the SNSPDs. In the single pump situations, the degenerate SFWM occurs in the quantum light source, generating photon pairs with pairwise correlation that are symmetric about the pump frequency, as schematically depicted in Fig. 3(b). We tune the pump frequency to correspond to 100-GHz ITU channels of C30, C40, and C50, and the obtained JSI results are shown in Fig. 3(c)-(e), respectively. It reveals the strong frequency anticorrelation determined by the energy conservation conditions and the tunability of the two-photon correlation. Under different pump frequencies, the coincidence line shifts parallel to the antidiagonal because the sum of the signal and idler frequencies varies with the pump frequency. The central point of the coincidence line is excluded from the test since it involves photons of the same frequency as the pump light. The corresponding coincidence to accidental coincidence ratios (CARs) for each frequency pair in three pump cases are given in Fig. 3(f)-(h). When approaching the pump frequency, the CAR decreases due to the influence of residual pump photons. For frequency pairs away from the pump frequency, the CAR is on the order of 1e3.

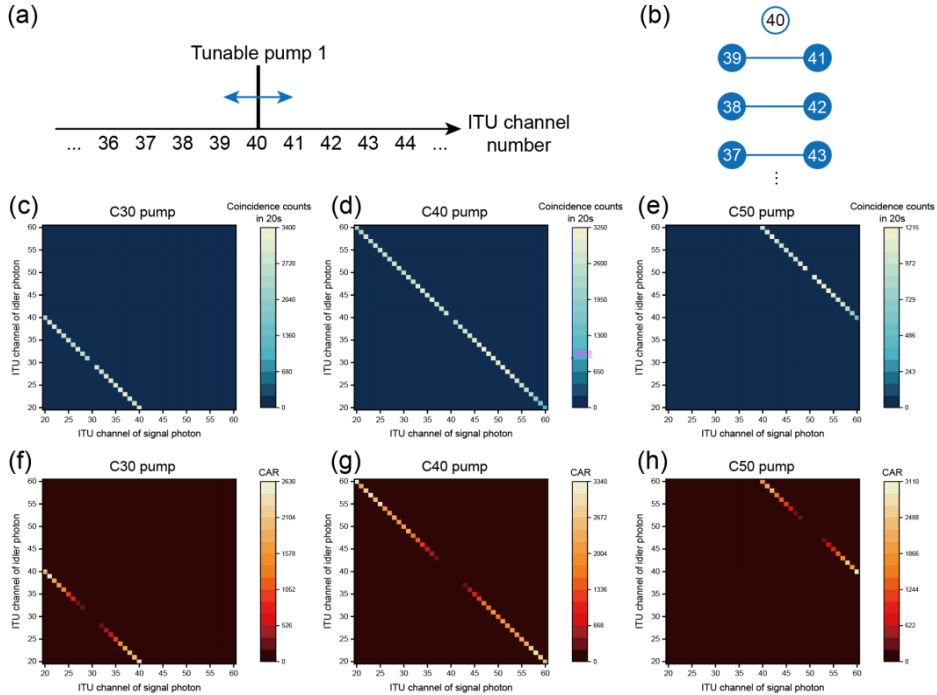

Fig. 3 Characterization of two-photon correlation in the case of a single pump. Schematic diagrams of (a) the pumping scheme using a tunable pump and (b) the generated quantum correlation between photons with different frequencies. The numbers correspond to the 100-GHz ITU channels. Assume that the pump frequency channel is C40, shown in hollow circles, the signal and idler channels are symmetric about the pump channel, shown in solid circles. Then the solid lines connecting the channels represent quantum correlations between entangled photons. (c)-(e)

The measured two-photon JSI under different pump frequencies. There is a coincidence line parallel to the antidiagonal in each pump condition, which shows the spectral anticorrelation determined by energy conservation. The photons at the pump frequency are excluded from the test. The coincidence window is 200 ps in the JSI measurements of this work. The integration time for each coincidence measurement is 20 s. (f)-(h) The CAR results under different pump frequencies.

Adding another pump laser to the system, a more complex two-photon correlation feature will emerge. The pumping scheme is shown in Fig. 4(a), and each pump power before coupling into the chip is about 2 mW. The degenerate SFWM process associated with each pump and the non-degenerate SFWM process associated with two pumps occur simultaneously, leading to a complex quantum correlation structure (see Fig. 4(b)). It can be divided into two sets of frequency channels with quantum correlations which can be viewed as two 2D square lattice configurations. Tuning the pump frequencies of the two pump lights, the obtained JSI results are shown in Fig. 4(c)-(e). Three coincidence lines emerge, with the middle line corresponding to the non-degenerate case and the other lines corresponding to the degenerate cases. It can be seen that the number of coincidence counts in the non-degenerate case is about 4 times higher than that in the degenerate cases, since the SFWM with distinct pump photons are more efficient in generating photon pairs than the degenerate SFWM when two pumps are launched with equal powers[24]. As the frequency difference of the pump lights increases, the gap between the three coincidence lines also increases, which is determined by energy conservation. Furthermore, in the dual pump configuration, it is found that strong lights are generated at specific frequencies. Assume that the pump frequencies are denoted by $\omega_1$ and $\omega_2$, the strong lights are generated at $2\omega_1 - \omega_2$ and $2\omega_2 - \omega_1$ due to the stimulated FWM process[29] where each pump laser can also be regarded as an input signal beam. Hence, the photons at these frequencies as well as those at the pump frequencies are excluded in the test. The corresponding CAR results for each frequency pair are given in Fig. 4(f)-(h). Similarly, the CAR in the non-degenerate case is also about 4 times higher than that in the degenerate cases. The reason is that under a fixed pump condition, the single-side count rate of each user is at the same order of magnitude, leading to similar accidental coincidence rates for all user links. Then, the difference in CAR is primarily attributed to the difference in the coincidence count rate.

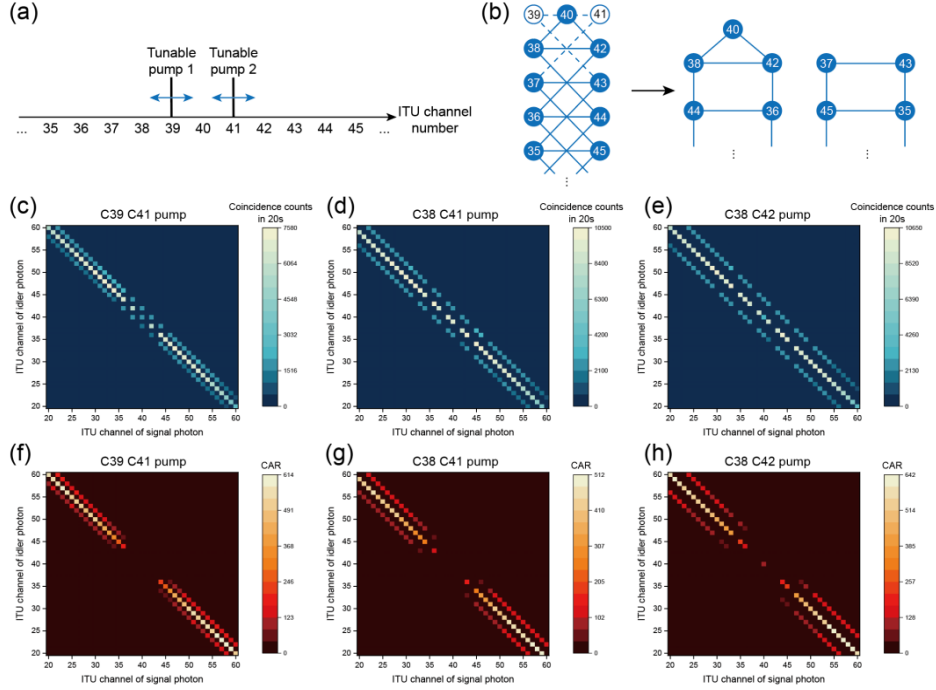

Fig. 4 Characterization of two-photon correlation in the case of dual pumps. Schematic diagrams of (a) the pumping scheme using two tunable pumps and (b) the generated quantum correlation between photons with different frequencies. The pump frequencies are assumed to be C39 and C41 for example. The dashed lines represent quantum correlations with undetected photons in our test due to pump lights and stimulated FWM. It can be divided into two sets of frequency channels with quantum correlations of 2D square lattice configurations. (c)-(e) The measured two-photon JSI under different pump frequencies. There are three coincidence lines parallel to the antidiagonal in each pump condition, which shows the spectral anticorrelation determined by energy conservation of degenerate and non-degenerate SFWM processes. The photons at the same frequencies as the generated strong light and the pump lights are excluded from the test. The integration time for each coincidence measurement is 20 s. (f)-(h) The CAR results under different pump frequencies.

Furthermore, the triple pump configuration enables more SFWM processes to occur simultaneously. The pumping scheme is shown in Fig. 5(a), and each pump power before coupling into the chip is about 2 mW as well. In this case, the degenerate SFWM process associated with each pump and the non-degenerate SFWM process associated with every two pumps occur simultaneously. The structure of the generated quantum correlation is more complex, which can be divided into two sets of 3D cubic lattices with different frequency channels (see Fig. 5(b)). There are six types of SFWM energy conservation relations based on the three pump frequencies. Therefore, the number of coincidence lines in the two-photon JSI is five or six (see Fig. 5(c)-(f)) depending on whether the pumps are equispaced in frequency or not. To avoid saturation of SNSPDs, there is an additional loss introduced by variable optical attenuators before detection of 3.7 dB, 7.5 dB, 4.6 dB, and 0 dB for Fig. 5(c)-(f), respectively. In addition to the phenomena discussed in the two configurations above, strong lights are generated at more specific frequencies in the triple pump configuration. Assume that the pump frequencies are denoted by $\omega_1$, $\omega_2$ and $\omega_3$, the strong lights are generated at $2\omega_1 - \omega_2$, $2\omega_1 - \omega_3$, $2\omega_2 - \omega_1$, $2\omega_2 - \omega_3$, $2\omega_3 - \omega_1$ and $2\omega_3 - \omega_2$ due to the stimulated FWM process. Given that FWM Bragg scattering involves two pump fields and a signal field[30], in the triple pump configuration, the strong lights are also generated at $\omega_1 + \omega_2 - \omega_3$, $\omega_1 - \omega_2 + \omega_3$ and $-\omega_1 + \omega_2 + \omega_3$. Similarly, the photons at these frequencies as well as the pump frequencies are discarded. The corresponding CAR results for each frequency pair are given in Fig. 5(g)-(j). The situations with equispaced pumps outperform the situation with non-equispaced pumps in terms of CAR, since the number of frequencies that generate strong light is higher in the latter case. As a result, the requirement for filtering will be higher.

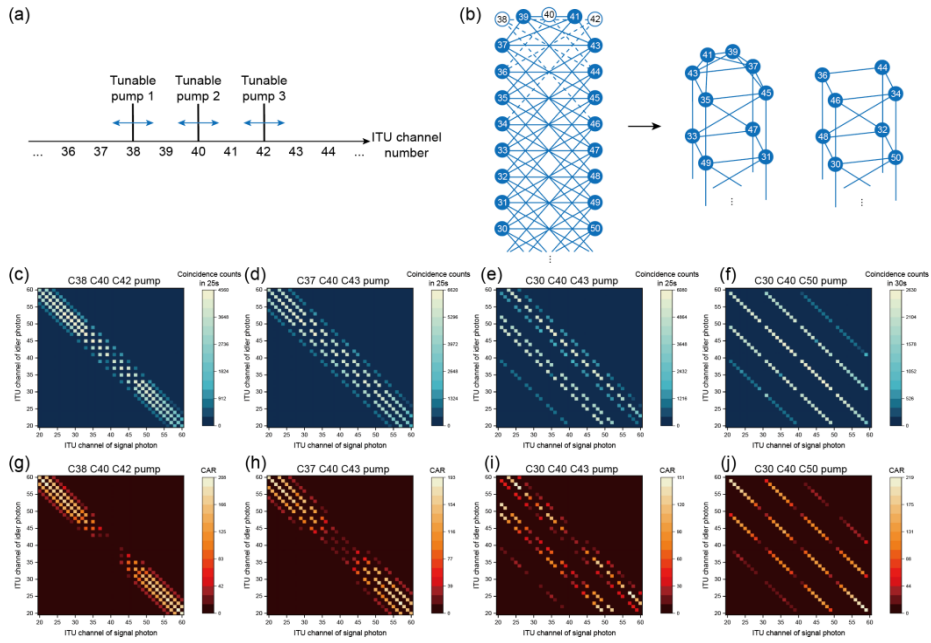

Fig. 5 Characterization of two-photon correlation in the case of triple pumps. Schematic diagrams of (a) the pumping scheme using three tunable pumps and (b) the generated quantum correlation between photons with different frequencies. The pump frequencies are assumed to be C38, C40, and C42 for example. It can be divided into two sets of 3D cubic lattices. (c)-(f) The measured two-photon JSI under different pump frequencies. There are five (six) coincidence lines parallel to the antidiagonal when $\omega_1$ and $\omega_3$ are symmetric (asymmetric) about $\omega_2$. The photons at the same frequencies as the generated strong light and the pump lights are excluded from the test. The integration time for each coincidence measurement is 25 s or 30 s. (g)-(j) The CAR results under different pump frequencies.

In addition to the two-photon correlation, we characterize the energy-time entanglement of some specific frequency pairs by Franson-type interference[31]. Under the single pump with frequency C40, Franson-type interference is implemented with each frequency pair symmetric to the pump frequency. For other pump cases, we randomly select one specific frequency pair for each SFWM process to perform Franson-type interference (see details and results in the Supplementary Material).

### 3.2 Fully-connected QKD network

Subsequently, a 10-user fully-connected QKD network is realized using the setup shown in Fig. 2(b). With a limited number of available SNSPDs, we measure the QKD results in a pairwise way, that is, we select two users to measure at the same time while the network is established under each pump configuration. The user pairs that do not receive entangled photon pairs are unable to establish QKD, so these links are not measured. In each detection basis, 70% of the coincidence counts are used for key generation and the rest are used for security analysis. The high-dimensional encoding parameters are optimized for each two-user case to maximize the secure key rate (SKR) (see details in the Supplementary Materials)[32].

The pump lights are switched between four configurations to effectively reconfigure the network topology. The measurement time is 10 min for each user link to generate secure keys in the asymptotic regime. The asymptotic SKR values in four pump configurations are shown in Fig. 6(a)-(d), respectively. The resulting network topologies (see insets) follow the prior designs. There is a difference in the SKRs between different user pairs, which is mainly attributed to two factors. First of all, as mentioned above, the coincidence count rate in the non-degenerate SFWM cases is about 4 times higher than that in the degenerate cases. Second, under a fixed pump condition, the CAR of different SFWM processes has the same relation as the coincidence count rate. Then, it causes the DO-QKD encoding dimension to differ between user pairs after parameter optimization. After time-sharing, the overall SKR is the total number of keys divided by the total time, as shown in Fig. 6(e). It can be seen that every user link above the diagonal has a positive SKR with an average value of 122.2 bps, realizing a 10-user fully-connected QKD network. Considering the finite-size effect, we also successfully generate secure keys for all the user links, however, a longer measurement time is needed for some user links (see details in the Supplementary Materials).

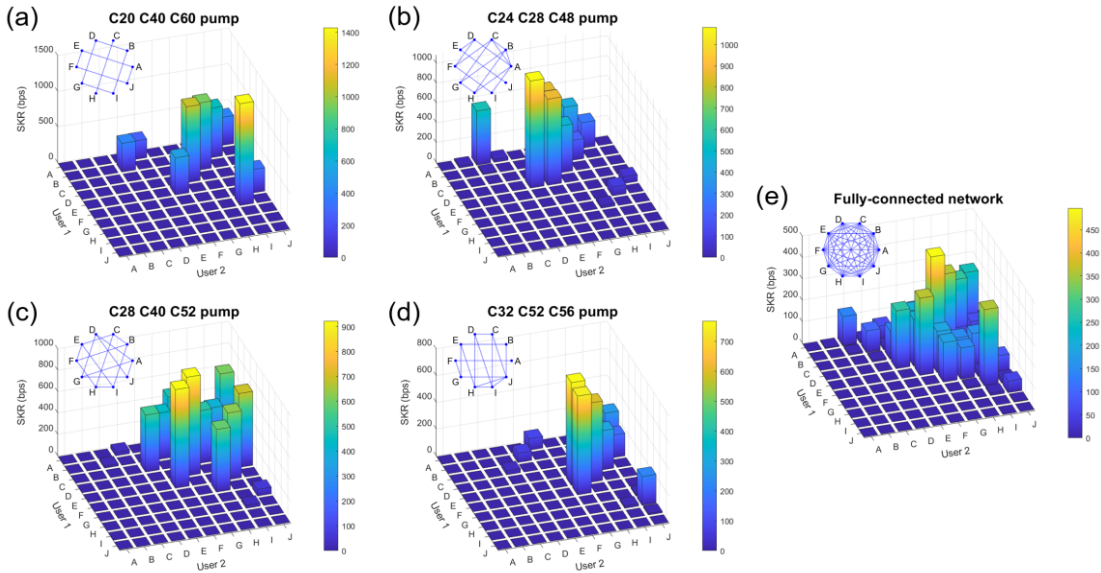

Fig. 6 Asymptotic SKR of the time-shared QKD network. (a)-(d) Temporary asymptotic SKR results under four pump configurations. Insets show the corresponding network topologies. (e) The overall asymptotic SKR result of the fully-connected network using time-sharing.

## 4 Discussion

We have demonstrated a reconfigurable entanglement distribution network scheme through pump management of an SFWM quantum light source and presented a 10-user fully-connected QKD network using time-sharing. The situations of a single pump, dual pumps, and triple pumps are investigated to show the tunability of the two-photon correlation generated by SFWM. There is a difference in the SKRs between user pairs, which is mainly due to the intrinsic nature of degenerate and non-degenerate SFWM processes. Fortunately, the tunability of pump frequencies brings the possibility to optimize the network performance with a preferred outcome by prior designing the pump configurations. Additional pump configurations can be introduced to compensate for the SKR difference. However, it may be at the cost of a decrease in the average SKR.

One of the advantages of our network scheme is that the frequency utilization is higher. It uses $N$ frequency channels to support an N-user fully-connected network. Compared with the $O(N^2)$ scaling of frequency channels in WDM-based fully-connected networks, the proposed scheme greatly relaxes the requirement for the number of frequency channels, thus it can be extended to more users. Using the ITU 100-GHz grid channels, our network scheme is expected to support up to 40 users with a series of designed pump configurations and excellent pump filtering. Even a larger-scale network with more than 100 users is possible by employing a finer frequency grid, e.g. the 50-GHz grid, or including the L-band frequencies.

In addition, another key feature of the proposed network is that each user link exclusively occupies a frequency pair of entanglement resource. There is no passive splitting in the distribution process of photons. On one side, the loss caused by the beam splitter is avoided. On the other side, it is possible to conduct multiple quantum information applications beyond QKD. In our experiment using multiple pumps, a user may be connected to more than one user, so it can still only support applications with post-processing, such as QKD. To distinguish each user pair, a single tunable pump is required to establish an entanglement distribution network where each user is connected to only one user. Hence, it can support quantum information applications without post-processing at the cost of using more pump configurations.

What is more, the pump management scheme offers reconfigurability to the entanglement distribution networks. The frequency and power of each pump can be easily tuned to reconfigure the logical topology of the network without changing the physical links. Compared with the reconfigurable network schemes using WSS[19-21] or q-ROADM[22, 23], the losses introduced by the optical switches and wavelength switches on the entangled photon pairs are avoided. Also, our scheme is not limited to the number of output ports of WSS or q-ROADM and has higher scalability.

The system loss is independent of the user number as well. We note that the network with pump management has limited reconfigurability and is unable to realize arbitrary topology at a time, so it is suitable for use with the time-sharing method.

The pump situation is not limited to triple pumps. Using more pump lights, even an optical frequency comb, to excite SFWM processes in the quantum light source may result in a more complex two-photon correlation feature, while the requirement for pump filtering and strong light filtering will become higher. If multiple lasers coherently pump the quantum light source, quantum optical frequency combs with different spectral correlation features can be generated, adding an extra degree of freedom for quantum information processing in the frequency domain[33-35]. The photons discussed above are in a bipartite entanglement state. When entangling multiple photon pairs, multiphoton entanglement would be generated with a high-dimensional lattice in the frequency domain, which would have great potential as a universal resource for measurement-based quantum computing[36].

## 5 Conclusion

In summary, we have proposed a reconfigurable entanglement distribution network scheme based on pump management and a time-sharing method. Single, dual, and triple pump lasers are used to excite the SFWM process in a silicon waveguide chip. When multiple pump lights excite the $\chi^{(3)}$ nonlinear material, SFWM processes with degenerate and non-degenerate pump photons occur simultaneously, forming complex two-photon correlation features that are characterized by JSI measurements. The energy-time entanglement of specific frequency pairs is verified by Franson-type interference. Further, we have realized a 10-user fully-connected QKD network under triple pumps as a benchmark application. The quantum light source for the network is pumped under four pump configurations and the network is constructed in a time-sharing way. Our results show the scalability, functionality, and reconfigurability of the proposed entanglement distribution network, which provides a potential architecture for future quantum networks.

**Acknowledgements. Acknowledgements.** This work was supported by the National Natural Science Foundation of China (92365210), National Key R&D Program of China (2018YFB2200400), Tsinghua Initiative Scientific Research Program and the project of JIAOT.

**Competing Interests.** The authors declare no competing interests.

**Author contributions.** W. Z. and J. L. proposed the scheme. J. L. and Z. L. performed simulations. J. L., D. L., and Z. L. performed experiments. Z. J designed and fabricated the silicon chip for the quantum light source. H. L. and L. Y. provided SNSPDs used in this work. W. Z. and J. L. wrote the manuscript. Y. H. revised the manuscript and supervised the project. X. F., F. L., and K. C. contributed to the experiment design and the revision of the manuscript.